\documentstyle[aps,prl,twocolumn,epsfig,floats,amssymb,amsfonts]{revtex} 

\begin{document} 
  
\draft 
 
\date{July 2001} 
 
\title{Microscopic versus mesoscopic
local density of states in one-dimensional localization} 
 
\author{H. Schomerus,$^1$ M. Titov,$^2$ P. W. Brouwer,$^3$ and C. W. J. Beenakker$^2$}
\address{$^1$ Max-Planck-Institut f\"ur Physik komplexer Systeme,
N\"othnitzer Str. 38, 01187 Dresden, Germany
}
\address{$^2$ Instituut-Lorentz, Universiteit Leiden, P. O. Box 9506,
2300 RA Leiden, The Netherlands
}
\address{$^3$
Laboratory of Atomic and Solid State Physics, Cornell University,
Ithaca, NY 14853-2501
}

\twocolumn[ 
\widetext 
\begin{@twocolumnfalse} 
\maketitle 

\begin{abstract} 
We calculate 
the probability
distribution of the local density of states $\nu$ in  a
disordered one-dimensional conductor or single-mode waveguide, attached at one end
to an electron or photon reservoir.
We show that  this distribution does not display a log-normal 
tail for small $\nu$, but diverges instead $\propto 
\nu^{-1/2}$. 
The log-normal  tail
appears if $\nu$ is averaged over rapid oscillations on the scale of the
wavelength. There is no such qualitative distinction between
microscopic and mesoscopic densities of states if the levels 
are broadened by inelastic scattering or absorption,
rather than by coupling to a reservoir.
\end{abstract} 
\pacs{PACS numbers: 72.15.Rn, 42.25.Dd, 73.63.Nm } 
\end{@twocolumnfalse} 
] 
\narrowtext 
 
Localization of wave functions by disorder can be seen in the
fluctuations of the density of
states, provided the system is probed on a sufficiently short length
scale \cite{EfetovBook,MirlinRep}. The local density
of states (LDOS) of electrons can be probed using the tunnel
resistance of a point contact \cite{Schmidt}
or the Knight shift in nuclear magnetic
resonance \cite{Fritschij}, 
while the LDOS of photons determines the rate of spontaneous
emission from an atomic transition \cite{Vries}. 
(In the photonic case one can study the effects of 
localization independently from those of interactions.)

For each length scale $\delta$ characteristic for the resolution 
of the probe, one can introduce a corresponding LDOS
$\nu_{\delta}$. It is necessary that $\delta$ is less than the
localization length, in order to be able to see the effects of
localization --- the hallmark \cite{AKL} 
being the appearance of logarithmically
normal tails $\propto\exp(-{\rm constant}\times\ln^{2}\nu_{\delta})$ in the
probability distribution $P(\nu_{\delta})$.

Much of our present understanding \cite{FM}
of this problem in a wire geometry
builds on the one-dimensional (1D) solution 
of Altshuler and Prigodin \cite{A&P}. 
In the simplest case one has a single-mode wire
which is closed at one end and attached at the other end to an electron
reservoir. (The optical analogue would be a single-mode wave\-guide that
can radiate into free space from one end.) In 1D the localization length
equals twice the mean free path $l$, which is assumed to be large compared to
the wavelength $\lambda$. One can then distinguish the microscopic LDOS
$\nu=\nu_{\delta}$ for $\delta\ll\lambda$, and the mesoscopic LDOS
$\tilde{\nu}=\nu_{\delta}$ for $\lambda \ll\delta\ll l$. While $\nu$
oscillates rapidly on the scale of the wavelength, $\tilde{\nu}$
only contains the slowly varying envelope of these oscillations.
Altshuler and Prigodin calculated the distribution $P(\tilde{\nu})$
and surmised that $P(\nu)$ would have the same log-normal tails.
We will demonstrate that this is not the case for the small-$\nu$
asymptotics.

The calculation of Ref.\ \cite{A&P} was based on the Berezinskii
diagram technique \cite{Berezinskii}, which reconstructs the probability
distribution from its moments. (An alternative approach \cite{McK}, 
using the method of supersymmetry, also proceeds via the moments.)
An altogether different scattering approach has been proposed by Gasparian,
Christen, and B\"uttiker \cite{GCB}, and more recently by 
Pustilnik \cite{Pustilnik}. We have pursued this approach and arrive at a
relation between $\nu$, $\tilde{\nu}$ and reflection
coefficients. This allows a direct calculation of the
distributions. We find that $P(\nu)$ and $P(\tilde{\nu})$ have the
same log-normal tail for large densities, but the asymptotics
for small $\nu$ and $\tilde{\nu}$ is completely different. The strong
fluctuations of $\nu$ on the scale of the wavelength lead to a
divergence $P(\nu) \propto \nu^{-1/2}$ for $\nu\to 0$, while the
distribution of the envelope vanishes, $P(\tilde{\nu}) \to 0$
for $\tilde{\nu}\to 0$. This qualitative difference between 
microscopic and mesoscopic LDOS is a feature of an open system.
Both $P(\nu)$ and $P(\tilde{\nu})$ vanish for small densities if
the wire is closed at both ends and the levels are broadened 
by inelastic scatterers (for electrons) or absorption
(for photons).

We consider a 1D wire and relate the microscopic
and mesoscopic LDOS at energy $E$ and at a  point $x=0$ 
to the reflection amplitudes $r_{\rm R}$, $r_{\rm L}$ 
from parts of the wire to the right and to the left of this point.
The Hamiltonian is $H=-(\hbar^2/2m)\partial^2/\partial x^2 +V(x)$
for non-interacting electrons. (For photons of a
single polarization we would consider the differential operator
of the scalar wave equation.) 
We will put $\hbar=1$ for convenience of notation. We start from the relation
between the LDOS and the retarded Green function,
\begin{eqnarray}
\label{definition}
&&\nu =-\pi^{-1} \,{\rm Im}\, G(0),\\
&&(E+i\eta-H)G(x) = \delta(x), 
\label{green}  
\end{eqnarray}
with $\eta$ a positive infinitesimal. We assume weak disorder
($k l\gg 1$, with $k=2 \pi/\lambda$ the wavenumber), so that 
we can expand the  Green function in scattering states 
in a small interval around $x=0$,
\begin{equation}
  G(x)\!=\! c_{\rm L} (e^{-i k x}\! +\! r_{\rm L} e^{i k x})\theta(-x) 
  \!+\! c_{\rm R} (e^{i k x}\! +\! r_{\rm R} e^{-i k x})\theta(x).
  \label{green2}  
\end{equation}
(The function $\theta(x)=1$ for $x>0$ and $0$ for $x<0$.)
The coefficients $c_{\rm L}$ and $c_{\rm R}$ are related by
the requirement that the Green function be continuous
at $x=0$,
\begin{equation}
  c_{\rm L}(1+r_{\rm L}) = c_{\rm R}(1+r_{\rm R}). 
  \label{green3} 
\end{equation}
Substitution of Eq.\ (\ref{green2}) into Eq.\ (\ref{green})
gives a second relation between $c_{\rm L}$ and $c_{\rm R}$,
from which we deduce that 
\begin{equation}
  G(0) = \frac{(1+r_{\rm L}) (1 + r_{\rm R})}{i v (1 - r_{\rm R} r_{\rm L})}, 
\end{equation}
with $v$ the velocity.
Using Eq.\ (\ref{definition}) we arrive at 
the key relation between the microscopic LDOS and 
the reflection coefficients,
\begin{equation}
\nu=(\pi v)^{-1}\,{\rm Re}\, (1+r_{\rm L})(1-r_{\rm 
R}r_{\rm L})^{-1}(1+r_{\rm R}).
\label{nu} 
\end{equation}

In order to perform the 
local spatial average that gives the mesoscopic 
LDOS $\tilde \nu$, 
we use that the reflection coefficients
oscillate on the scale of the wavelength.
If we shift $x_0$ slightly away from the origin to 
a point $x'$, one has
$r_{\rm L}\to e^{2ik x'}r_{\rm L}$ and
$r_{\rm R}\to e^{-2ik x'}r_{\rm R}$. 
The product $r_{\rm R}r_{\rm L}$, however, does not display these 
oscillations---only this combination should be retained. 
Hence
\begin{equation}
\tilde{\nu}=(\pi v)^{-1}\,{\rm Re}\, 
(1+r_{\rm R}r_{\rm L})\,(1-r_{\rm R}r_{\rm L})^{-1}. 
\label{tnu} 
\end{equation}
In what follows we will measure $\nu$ and $\tilde{\nu}$
in units of $\nu_0=(\pi v)^{-1}$, which is the macroscopic density of
states and the ensemble average of $\nu$, $\tilde{\nu}$. 
 
Let us now demonstrate the power of the two simple relations
(\ref{nu})  and (\ref{tnu}).
We take the wire open at the left end and
study the density at a distance $L$ from this opening. 
At the right end the wire is assumed to be closed,
giving rise to a reflection coefficient
$r_{\rm R}=\exp(i \phi_{\rm R})$ with
uniformly distributed phase $\phi_{\rm R}$
in the interval $(0,2\pi)$.
The reflection coefficient  $r_{\rm L}=\sqrt{R}\exp(i\phi_{\rm L})$
is parameterized through the uniformly distributed phase
$\phi_{\rm L}$ and the reflection probability $R$ in the
interval $(0,1)$.
The ratio
\begin{equation}
\label{T} 
u=(1+R)\,(1-R)^{-1}
\end{equation}
has the probability distribution \cite{Abrikosov}
\begin{equation}
\label{normalx} 
\rho(u)=\frac{e^{-s/4} }{\sqrt{\pi} (2s)^{3/2}} 
\int\limits_{{\rm arcosh}\,u}^\infty dz\; \frac{z\; e^{-z^2/4
s}}{(\cosh{z}-u)^{1/2}},
\end{equation}
with $s=L/l$ and $l$ the mean free path for backscattering.
The mesoscopic LDOS (\ref{tnu}) can be written in terms of the 
variables $u$ and $\phi=\phi_{\rm L}+\phi_{\rm R}$,
\begin{equation}
\tilde \nu=\left(u-\sqrt{u^2-1}\cos \phi \right)^{-1}. 
\end{equation}
Averaging first over $\phi$ we find 
\begin{equation}
P_{\rm open}(\tilde\nu)=\frac{\tilde\nu^{-3/2}}{ \pi \sqrt{2} } 
\int\limits_{a}^\infty 
du\;\frac{\rho(u)}{\sqrt{u-a}}, \quad a=\case{1}{2}(\tilde\nu+{\tilde\nu}^{-1}).
\label{general} 
\end{equation}
The subsequent integration with Eq.\ (\ref{normalx}) yields 
\begin{equation}
P_{\rm open}(\tilde \nu)=\frac{\tilde{\nu}^{-3/2} e^{-s/4}}
{2\sqrt{\pi s}}
\exp\left(-\frac{1}{4s}\ln^2
\tilde\nu \right).
\label{mdosopen}
\end{equation}

The distribution function (\ref{mdosopen})
is the celebrated result of Altshuler and Prigodin \cite{A&P}.
It displays log-normal tails for both large and small 
values of $\tilde \nu$. Indeed, the two tails are linked by
the functional relation \cite{FM}
\begin{equation}
P(1/\tilde\nu)={\tilde \nu}^3 P(\tilde\nu).
\label{eq:func}
\end{equation}
This relation follows directly from  Eq.\ (\ref{general}) and hence
requires only a uniformly distributed phase $\phi$, 
regardless of the distribution function $\rho(u)$
of the reflection probability. As we will now show, 
such a relation does not hold in general for the
microscopic LDOS $\nu$, and
the asymptotics of its distribution function 
for small and large values of $\nu$ can be entirely different.

The calculation is facilitated by the fact that $\nu$
is related to $\tilde\nu$ by 
\begin{equation}
\label{simple}
\nu=2 \tilde\nu \cos^2(\phi_{\rm R}/2)
\qquad \mbox{if}\; |r_{\rm R}|=1. 
\end{equation}
Moreover, $\tilde\nu$ is statistically 
independent of $\phi_{\rm R}$ because the latter enters $\tilde\nu$ only 
in combination with $\phi_{\rm L}$, which itself is uniformly distributed. 
The distribution of the microscopic LDOS
hence follows directly from Eq.\ (\ref{mdosopen}),
\begin{equation}
P_{\rm open}(\nu)=\frac{\nu^{-3/2}e^{-s/4}}{\pi\sqrt{2 \pi s}}
\int_0^1\frac{dt}{\sqrt{1-t}} 
\exp\left(-\frac{1}{4 s} \ln^2 \nu \right),
\label{form3} 
\end{equation}
where we substituted $t=\cos^2(\phi_{\rm R}/2)$. 
The asymptotic behavior is
\begin{mathletters}
\begin{eqnarray}
&&P_{\rm open}(\nu)=\frac{\exp(3s/4)}{2^{1/2}\pi} \; \nu^{-1/2}, 
\qquad \nu \ll e^{-s},\\ 
&&P_{\rm open}(\nu)=\frac{2^{1/2}\exp(-s/4)}{s^{1/2}\pi^{3/2}}
\nu^{-3/2}, 
\quad e^{-s} \ll \nu \ll e^{s},\\ 
&&P_{\rm open}(\nu)= 
\frac{\exp[-s/4-\ln^2(\nu/2)/4s]}
{\pi\nu^{3/2}\ln^{1/2}(\nu/2)},
\qquad \nu \gg e^{s}. 
\end{eqnarray}
\end{mathletters}
In the second and third region this is similar to the behavior of
$P_{\rm open}(\tilde\nu)$ in Eq.\ (\ref{mdosopen}).
In the region of the smallest densities, however,  
$P_{\rm open}(\nu)$ is not log-normal like $P_{\rm open}(\tilde{\nu})$
but diverges $\propto \nu^{-1/2}$.

The different tails arise
from two qualitatively different mechanisms
that produce small values of $\nu$ and $\tilde\nu$.
For the mesoscopic LDOS this requires remoteness 
of $E$ from the eigenvalues of wave functions
localized within a localization length around $x_0$.
As a consequence, $P(\tilde\nu)$ is intimately linked to the
distribution function of resonance widths \cite{MirlinRep}.
Small values of the microscopic LDOS
$\nu$ are attained at nodes of the wave function
which solves the wave equation with open boundary conditions,
independent of the energy.
The nodes are completely determined by the
small-scale structure of the wave function, which is 
a real standing wave $\propto \cos{(k x+\alpha)}$ with
random phase $\alpha$ \cite{FM}. [We recognize the square
of this wave amplitude in Eq.\ (\ref{simple}).]
The resulting $\nu^{-1/2}$ divergence of the probability 
distribution has the same origin as in the Porter-Thomas
distribution for chaotic wave functions \cite{Mehta}.
 
\begin{figure} 
\epsfxsize\columnwidth 
\centerline{\epsffile{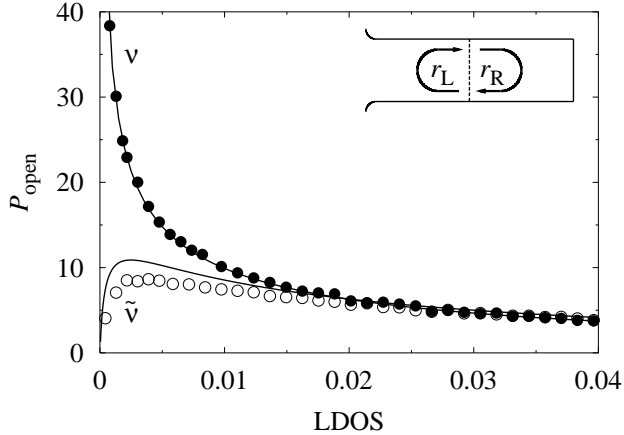}} 
\medskip 
\caption{Distributions of the microscopic 
local density of states (LDOS) $\nu$
and the mesoscopic LDOS $\tilde{\nu}$
for the open wire
at a distance  $L=2\, l$ from the opening. 
(Both are measured in units of their mean
$\nu_0=(\pi v)^{-1}$.)
Solid curves 
are given by Eqs.\ (\ref{mdosopen}) and (\ref{form3}). 
The data points 
result from a numerical simulation for a wire 
of length $10\, l$ with no adjustable parameter.
The inset shows the geometry of the open wire
(not to scale).
} 
\label{fig1} 
\end{figure} 

The two distributions for the open wire 
are plotted in Fig.\ \ref{fig1}, 
together with the result of a numerical simulation in which the Green 
function inside the wire is calculated recursively \cite{recgf}. 
The comparison of theory and numerics
is free of any adjustable parameter---the velocity was taken from the
dispersion relation, and the mean free path was obtained 
from the disorder strength within the Born approximation.
 
We now show that this qualitative difference 
between the microscopic 
and mesoscopic LDOS is absent in a closed wire.
If the wire is decoupled from the reservoir
we need another source of level broadening to regularize the 
delta functions in the LDOS.  Following Ref.\ \cite{A&P}, we
will retain a finite imaginary part $\eta$ of the energy,
corresponding to spatially uniform absorption (for photons) or
inelastic scattering (for electrons), with rate $2\eta $.
Eqs.\ (\ref{nu}) and (\ref{tnu}) still hold provided $\eta \ll E$.
The reflection coefficients can be written as 
\begin{equation}
r_{\rm R,L}=\sqrt{R_{\rm R,L}}e^{i\phi_{\rm R,L}}, 
\end{equation}
where $\phi_{\rm R}$ and $\phi_{\rm L}$ are uniformly distributed phases 
and $R_{\rm R}$, $R_{\rm L}$ are independent  
reflection probabilities. In an infinitely long wire they
have the same distribution \cite{Kumar}
\begin{equation}
\label{R} 
\rho(R)=\frac{\omega\, e^{\omega}}{ (1-R)^2}\exp{\left[ 
-{\omega (1-R)^{-1}}\right]},
\quad \omega=4 \eta l/v. 
\end{equation}

\begin{figure} 
\epsfxsize\columnwidth 
\centerline{\epsffile{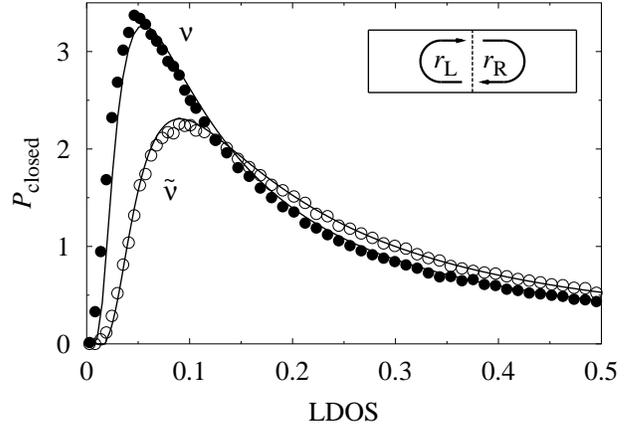}} 
\medskip 
\caption{Same as in Fig.\ \ref{fig1} but for the closed wire
with dimensionless absorption rate $\omega=1/6$.  Solid curves 
are given by Eqs.\ (\ref{fullform}) and (\ref{ldosabs}). The data points 
result from a numerical simulation for a wire 
of length $55\, l$, with the LDOS computed halfway in the wire.}
\label{fig2} 
\end{figure} 

After elimination of the phases the distribution
of the mesoscopic LDOS takes again the form
(\ref{general}), where $u$ now stands for the combination
\begin{equation}
u=(1+R_{\rm R} R_{\rm L})\,(1-R_{\rm R} R_{\rm L})^{-1}.
\end{equation}
Eq.\ (\ref{R}) implies for $u$ the distribution
\begin{equation}
\label{exact} 
\rho(u)=\omega^2\left(1-\frac{\partial }{\partial \omega}\right)
e^{-\omega(u-1)} 
K_0\left(\omega\sqrt{u^2-1}\right).
\end{equation}
The resulting distribution function of the mesoscopic LDOS is
\begin{eqnarray}
&&P_{\rm closed}(\tilde\nu)=\frac{\omega^2 \tilde{\nu}^{-3/2}
}{ \pi \sqrt{2} }
\int_a^\infty
\frac{du}{\sqrt{u-a}}\; e^{-\omega(u-1)}\nonumber \\
&&\;\times \left[
u K_0\left(\omega \sqrt{u^2-1} \right)
\!+\!\sqrt{u^2-1}K_1\left(\omega\sqrt{u^2-1}\right)
\right],
\label{fullform}
\end{eqnarray}
with $a$ defined in Eq.\ (\ref{general}). It
vanishes for small densities as
\begin{equation}
P_{\rm closed}(\tilde\nu)= 2^{-1/2}\omega {\tilde \nu}^{-2}
\exp(-\omega/\tilde\nu),\qquad  \tilde{\nu} \ll \omega.
\label{mdosabs} 
\end{equation}
This should be compared with the known distribution \cite{A&P}
\begin{equation}
P_{\rm
closed}(\nu)=\sqrt{\frac{2\omega }{ \pi}} \nu^{-3/2}\exp
\left[\omega-\case{1}{2}\omega(\nu+\nu^{-1})\right]
\label{ldosabs}
\end{equation}
of the microscopic LDOS.
In contrast to the open wire, both distributions 
vanish for $\nu$, $\tilde\nu\to 0$.
This is illustrated in Fig.\ \ref{fig2},
which compares the analytical predictions to numerical data
obtained by diagonalization of a Hamiltonian.
The comparison is again free of any adjustable parameter.

We note in passing that the asymptotic behavior
(\ref{mdosabs}) differs from the
asymptotic behavior
\begin{equation}
P_{\rm closed}(\tilde\nu)\neq  \case{1}{4}(\pi\omega)^{1/2} 
{\tilde\nu}^{-3/2}\exp(-\pi^2\omega/16\tilde\nu),
\label{apold} 
\end{equation}
given in Ref.\ \cite{A&P} for $\omega\ll 1$. 
There the distribution function was reconstructed from
the leading asymptotics of the moments 
$\lim_{\omega\to 0} \langle {\tilde\nu}^n\rangle= 
\omega^{1-n} n!/(2n-1)$.
This would be a valid procedure if the
distribution would depend only on the product
$\omega \tilde{\nu}$ in the limit $\omega \to 0$,
which it does not. The subleading terms of
the moments have to be included for $\tilde{\nu} \lesssim \omega$.
Indeed, our distribution function has the same leading
asymptotics of the moments, but has a different functional form.
This illustrates the potential pitfalls
of the restoration procedure
which are circumvented by our direct method.

In conclusion, we have given exact results for the distributions of the 
local densities of states in one-dimensional localization,
contrasting the microscopic length scale (below the wavelength)
and mesoscopic length scale (between the wavelength and the 
mean free path). Contrary to expectations in the literature,
the log-normal asymptotics at small densities applies 
only to the mesoscopic LDOS $\tilde{\nu}$, while
the distribution of the microscopic LDOS $\nu$ diverges 
$\propto \nu^{-1/2}$ for $\nu \to 0$.
This is of physical significance because many of the local probes 
act on atomic degrees of freedom and hence measure $\nu$ 
rather than $\tilde{\nu}$. The strong length scale dependence
of the LDOS disappears if the electrons (or photons)
are scattered inelastically (or absorbed) before reaching the
reservoir. Both $P(\nu)$ and $P(\tilde{\nu})$ then have an exponential cutoff
at small densities. 

It is an interesting open problem whether
the qualitative distinction between $\nu$ and $\tilde\nu$ 
in an open wire carries
over to the quasi-one dimensional geometry with $N>1$ modes.
An analytic theory could build on the multi-channel generalization of
Eq.\ (\ref{nu}),
\begin{equation}
\nu=\,{\rm Re}\,{\rm Tr}\, {\hat M}  (1+{\hat r}_{\rm L})(1-{\hat r}_{\rm R} 
{\hat r}_{\rm L})^{-1}(1+{\hat r}_{\rm R}).
\end{equation}
Now ${\hat r}_{\rm L}$ and ${\hat r}_{\rm R}$ are $N\times N$ 
reflection matrices 
and the matrix ${\hat M}_{nm}\!=\!2 (\pi A)^{-1} (v_nv_m)^{-1/2} 
\sin(\vec{q}_n\cdot\vec{r}_0) 
\sin(\vec{q}_m\cdot\vec{r}_0)$
contains the weights of the $N$ scattering states
with transversal momentum $\vec{q}_n$ and longitudinal velocity 
$v_n$ at the transversal position $\vec{r}_0$ on the cross section
of the wire (area $A$).

Another promising direction for future research
is to study what happens to the LDOS if the wire is coupled
to a superconductor rather than to a normal electron reservoir
\cite{Devoret}. The convenient expressions for $\nu$ and $\tilde{\nu}$
in terms of the reflection matrices from two independent parts of
the wire, derived in this paper, can be directly generalized to
include Andreev reflection at the interface.

We acknowledge discussions with A.~D.~Mirlin and M.~Pustilnik.
This work was supported by the Dutch Science Foundation NWO/FOM,
by the NSF under grant no. DMR 0086509, and by the Sloan foundation.

\end{document}